\documentstyle[12pt]{article}
\input{psfig}
\def\fun#1#2{\lower3.6pt\vbox{\baselineskip0pt\lineskip.9pt
  \ialign{$\mathsurround=0pt#1\hfil##\hfil$\crcr#2\crcr\sim\crcr}}}
\def\reff{\vskip 0.5cm\par\noindent\hangafter=1\hangindent=1cm} 
\def\mathrelfun#1#2{\lower3.6pt\vbox{\baselineskip0pt\lineskip.9pt
  \ialign{$\mathsurround=0pt#1\hfil##\hfil$\crcr#2\crcr\sim\crcr}}}

\def\eV  {{\rm \hbox{e\kern-0.14em V}}}
\def\keV {{\rm \hbox{ke\kern-0.14em V}}}
\def\MeV {{\rm \hbox{Me\kern-0.14em V}}}
\def\GeV {{\rm \hbox{Ge\kern-0.14em V}}}
\def\test{\theta^{\rm cmb}}
\def\ttrue{t^{\rm cmb}}
\def\sigz{\sigma_\theta^{(0)}}
\def\sigzt{\sigma_\theta^{(0)^2}}
\def\sigsh{\sigma_{\rm shape}}
\def\fdf{\rm FDF}
\def\tobs{T}
\def\nch{N_{\rm ch}}
\def\nf{N_{\rm fg}}

\def\sn{\sigma}
\def\nc{C}

\def\fcmb{F^{0}}
\def\ssp{{\cal F}}
\def\zr{z^{(r)}}
\def\zs{z^{(s)}}
\def\perp#1{\vec #1_\bot}
\def\ben{\begin{equation}}
\def\ee{\end{equation}}
\def\be#1{\begin{equation}\label{#1}}

\def\ff{\vec F^1 \cdot \vec F^1}
\def\fc{\hat F^0 \cdot \vec F^1}
\begin{document}
\baselineskip=15pt
\pagestyle{empty}
\begin{center}
\bigskip

\rightline{FERMILAB--Pub--95/375-A}
\rightline{submitted to {\it Astrophysical Journal}}

\vspace{.2in}
{\Large \bf Determining cosmic microwave background anisotropies
in the presence of foregrounds}
\bigskip

\vspace{.2in}
Scott Dodelson\\
\vspace{.2in}
{\it NASA/Fermilab Astrophysics Center\\
Fermi National Accelerator Laboratory, Batavia, IL~~60510-0500}\\

\end{center}

\vspace{.3in}

\centerline{\bf ABSTRACT}
\bigskip
Separating foregrounds from the signal is one of
the big challenges in cosmic microwave background (CMB)
experiments. A simple way to estimate
the CMB temperature in a given pixel
is to fit for the amplitudes of the CMB and the
various foreground components.
The variance squared  of this estimator is shown
to be equal to $[(\fdf)^2\ \sigzt + \sigsh^2]$, where $\sigz$
is the variance in the absence of foregrounds; $\sigsh$ is the variance
due to the uncertainty in the shapes of the foreground components;
and \fdf\ is the {\it foreground degradation factor}. This one number,
the \fdf,
gives a good indication of the ability of a given experiment
to disentangle the CMB from foreground sources. A variety of applications
relating to the planning and analyzing of experiments is presented.

\newpage
\pagestyle{plain}
\setcounter{page}{1}
\newpage

\section{Introduction}

The cosmic microwave background
encodes a great deal of information about our universe. In particular
the anisotropies -- and especially those on small scales --
 are sensitive to many cosmological parameters and to
the initial perturbations which grew into the large
structures observed today.  Thus a map of the anisotropies
in the CMB on small scales can unequivocably answer
questions that have plagued cosmologists
for decades (or longer). For this reason, a number of
groups have set out to make such maps of the sky at varying
angular resolution, typically better than half a degree.

There are many complicated experimental issues involved in making such maps.
However, even an experiment perfectly designed to
minimize atmospheric contamination, sidelobes, $1/f$ noise, etc.
still has to deal with the reality of the sky. And this reality includes
not only the ``signal'' in the form of CMB anisotropies but
also ``noise'' in the form of galactic and extragalactic foregrounds.
The most powerful way to extract the CMB signal from foreground
contamination is to take measurements at many different frequencies.
The CMB anisotropies vary with frequency differently than do the
foregrounds. By using the knowledge we have about these different
spectral shapes, we can conceivably extract the CMB component from
the total signal.

In this paper, I will discuss how to perform this extraction.
For a given set of frequencies and given number of foregrounds one
wants to eliminate, we can define an estimator, $\test$,
for the
true CMB temperature by
fitting the amplitudes of a CMB component and various
foreground components to the observed temperatures in each channel.
On average this estimator will equal the true
CMB temperature $\ttrue$. The variance of the estimator depends
on the instrumental and atmospheric noise of course. But it also
depends on the frequency coverage and the foregrounds. In fact
 the variance
can be simply expressed as
\begin{equation}
\sigma_\theta^2
= (\fdf)^2\ \sigzt + \sigsh^2,
\label{finalvar}
\end{equation}
where $\sigz$ is the variance in the absence of foregrounds
(due to instrumental and atmospheric noise) and $\sigsh$ is the
contribution to the variance due to the uncertainty in the
spectral shapes of the foregrounds. In many cases, $\sigsh$
will be small, so
the variance is enhanced over the
no-foreground case by the {\it foreground degradation factor}, $\fdf$.
By construction, \fdf\ is greater than or equal to one. Thus the
effectiveness of any given
set of frequencies can be expressed by this one number. If the \fdf\
for a given frequency set is large, then the contaminating
foreground is troubling; if \fdf\ is close to one, then foregrounds
may be effectively eliminated.

Another feature of equation \ref{finalvar}\ deserves mention.
The first term is proportional to the instrumental and atmospheric
noise; we will see that the second is proportional to the rms
amplitude of the foregrounds. While typically the first term dominates,
there are situations -- e.g. in experiments with very low noise
per pixel or experiments in dusty regions of the sky -- where the
second term is most important. In these cases, the channels should
be constructed to minimize $\sigsh$.

Section 2 presents a prescription for calculating
\fdf\ and $\sigsh$
for a given experimental configuration and set of foregrounds.
Some of the details are relegated to Appendix A.
Section 3 presents a number of applications of this analytic
technique; questions which
might come up in designing or analyzing an experiment which can
be simply addressed with the concept of \fdf\ and $\sigsh$.

This method of estimating the CMB temperature was carried
out by the MSAM team in Cheng et al (1994) when analyzing their
data. In several previous papers (Dodelson \& Stebbins 1994
and Dodelson \& Kosowsky 1995) my collaborators and I analyzed
a variety of experiments using an apparently different technique,
that of marginalization. In Appendix B, I show that the two
techniques are in fact identical.

I should point out that there has already been a good deal of work
on the issue of foregrounds. Perhaps the most influential has been
the paper by Brandt et al. (1994). Without getting into the details
of their work, I simply point out that their basic technique is the
Monte Carlo. Here I am more interested in seeing what can be done
analytically. In \S 3.4, I compare this analytic approach
with their Monte Carlo methods and find excellent agreement.
The work of Toffolatti et al (1994), Danese et al (1995), and
Tegmark
$\&$ Efstathiou (1995) uses information from other maps, such as
the IRAS (Neugebauer et al 1984) map of dust.
Although the formalism discussed in this paper
can probably be extended to include such maps, here I do not attempt
to do so. So the conclusions reached
here are probably on the conservative side (I assume that less is known about
the foregrounds). The fact that these conclusions are still reasonably
optimistic is encouraging and offers still more evidence that foregrounds
will {\it not} be an intractable problem for a satellite mission.

\section{CMB Estimator and Variance}

This section is divided into three parts. First there a brief discussion
of notation; this provides the information necessary to translate
the experimental/foreground
information into the vectors used to calculate \fdf\
and $\sigsh$. The second subsection
derives the estimator of the CMB temperature and its variance. One
simply performs a best fit to the free parameters: the amplitude of
the various components. Calculating the variance of this estimator
leads immediately to the concept and definition of \fdf\
and $\sigsh$.
Section 2.3 then presents a
simple formula for the \fdf\ in the presence of one and two foregrounds.

\subsection{Notation}

I will label the number of frequency channels with a subscript
$a=1,\dots,\nch$. The observed {\it antenna} temperature in
each channel is
denoted $\tobs_a$. It will be convenient to group
all $\nch$ of these numbers into an $\nch-$ dimensional vector
$\vec\tobs$.
The observed signal is composed
of the CMB component, foreground components, and noise, so
\begin{equation}
\vec\tobs = \sum_{i=0}^{\nf} \vec T^i  + \vec N
\label{obssig}
\end{equation}
where the CMB component has superscript $0$ and the
$i=1,\ldots,\nf$ foreground components are appropriately
superscripted; $\vec N$ denotes the contribution to the signal
from instrumental and atmospheric noise. The noise is
assumed Gaussian with
\begin{equation}
\langle \vec N \rangle = 0  \qquad\qquad; \qquad\qquad
\langle N_a N_b \rangle = \nc_{ab}.
\label{noiseapp}
\end{equation}
Throughout, $\vec\tobs^i$
will be used to refer to the true temperatures on the sky. These are
to be distinguished from the estimators,
$\vec \Theta^i$, which represent our best guess
about these temperatures. These estimators
will assume that the shape of the foregrounds and CMB are
known and take the amplitudes as free parameters. Thus, we set
\begin{equation}
\vec\Theta^i = \vec F^i \theta^i
\label{intsha}
\end{equation}
where $\theta^i$ is the (unknown) amplitude of the $i^{\rm th}$ component
and $\vec F^i$ is the (presumed) shape of that component.
As a simple
example consider the CMB component. We
know that it has a blackbody shape, after subtracting off
the mean,
\begin{equation}
\hat F^0_a = {1\over 2\nu_a^2}
\ {d\over dT} B_{\nu_a}(T) = {x_a^2 e^{x_a} \over
(e^{x_a} - 1)^2}
\label{deffo}
\end{equation}
where $x_a \equiv 2\pi \hbar \nu_a/k_B \bar T$ and $\bar T=2.726^\circ$K,
the average temperature. With this shape vector, the amplitude $\theta^0$
is the estimate of the {\it thermodynamic} temperature anisotropy,
which of course is frequency independent. Note that in the
Rayleigh-Jeans limit $(x_a\rightarrow 0)$, $\hat F^0_a \rightarrow 1$.
The CMB shape vector has a $\hat{}$ over it to
denote unit vector. That is, $\hat F^0\cdot \hat F^0 = 1$, where
the dot product of any two vectors is defined as
\begin{equation}
\vec T \cdot \vec S \equiv
\sigzt
\sum_{a,b=1}^{\nch} T_a {\nc^{-1}}_{ab} S_b .
\label{defdot}
\end{equation}
The prefactor here, $\sigzt$, is the variance
in the absence of foregrounds and can be written as
\begin{equation}
\sigzt \equiv {1\over \sum_{a=1}^{\nch}
\hat F^0_a \hat F^0_b {C^{-1}}_{ab}  }
\label{defsigz}
\end{equation}
We will see shortly that this is indeed the variance
in the no-foreground case, but one can immediately see
that this is reasonable by
considering the case of equal and uncorrelated noise with
variance $\sigma$ in each
channel in an experiment with frequencies
in the Rayleigh-Jeans limit. Then $\sigz \rightarrow \sigma/\sqrt\nch$,
the correct limit.
Finally, it will prove useful to introduce the
$(\nf+1)\times(\nf+1)$ matrix
\begin{equation}
K_{ij} \equiv \vec F^i \cdot \vec F^j.
\label{defk}
\end{equation}

\subsection{Best-Fit Estimator and its variance}

To determine the amplitudes $\theta^i$ of the various components,
we can minimize the variance:
\begin{equation}
{\partial\over \partial\theta^i} \langle\left(
        \vec\tobs - \sum_{i=0}^{\nf} \vec F^i \theta^i
\right)^2\rangle = 0.
\label{minvar}
\end{equation}
Appendix A provides the straightforward details of this minimization.
The result is that the estimator for the CMB temperature is
\begin{equation}
\theta^0 = \sum_{j=0}^{\nf} {K^{-1}}_{0j} \vec F^j \cdot \vec\tobs.
\label{eqest}
\end{equation}
It is important to note that the estimator
in equation \ref{eqest}\ is linear in the observed temperature
$\vec\tobs$. Therefore, if the noise around $\vec\tobs$ is Gaussian,
then the noise around $\theta$ will also be Gaussian. Had we allowed the
foreground shapes to vary as well, the transformation would no longer
be linear, and there would be no reason to expect the noise
to be Gaussian.

Equation \ref{eqest}\
is an estimate for the thermodynamic temperature
anisotropy of the CMB. How good an estimator is it?
To answer this question, we need to compute
\begin{equation}
\sigma_\theta^2 \equiv
 \langle\left( \theta^{0} - t^{0} \right)^2 \rangle
\label{defvar}
\end{equation}
where $t^{0}$ is the true CMB thermodynamic temperature
on the sky.
A short calculation (presented in Appendix A)
shows that this variance is given by equation \ref{finalvar}
with
\begin{equation}
\fdf \equiv \sqrt{ {K^{-1}}_{00} }
\label{deffdf}
\end{equation}
and:
\begin{equation}
\sigsh^2 \equiv
\left(
\sum_{i=1}^{\nf} \vec T^i \cdot \sum_{j=0}^{\nf} {K^{-1}}_{0j} \vec F^j
\right)^2.
\label{sigfg}
\end{equation}

One important limit of equation \ref{finalvar}\ is when no foregrounds
are projected out $\nf=0$. In that case, the matrix $K$ has only one
component, the ${}_{00}$ component which is unity. Thus $\fdf=1$
and the variance is equal to $\sigz$, as defined in equation
\ref{defsigz}. Another important limit is when
the foreground shapes are known. In this limit, $\sigsh$ vanishes.
To see
this, note that if we have chosen the correct $\vec F^i$ for the foregrounds,
then the true foregrounds are proportional to $\vec F^i$. Then,
the dot product $T^i \cdot \vec F^j$ in
equation \ref{sigfg}\ is proportional to $K_{ij}$. When
multiplied by ${K^{-1}}_{0j}$
and summed over $j$, this gives a delta function, $\delta_{0i}$
which vanishes for all foregrounds $i>0$.

\subsection{{\fdf} in the presence of one or two foregrounds}

The $\fdf$ can be easily calculated via equation \ref{deffdf}\
once the matrix $K$ is known. $K$ in turn depends
on the assumed foreground shapes via equation \ref{defk}.
Here I present the results for
$\fdf$ in the cases where (i) one foreground is to be projected
out and (ii) two foregrounds are to be removed.

If there is one foreground to be removed with shape vector
$\vec F^1$, then the matrix $K$ depends on only the dot
products $\ff$ and $\fc$:
\begin{equation}
K = \left( \matrix{1 && \fc \cr
                 \fc && \ff } \right)
\label{koned}
\end{equation}
The inverse of this $K$ is readily obtained:
\begin{equation}
K^{-1} = {1\over \ff - (\fc)^2} \left( \matrix{\ff && -\fc \cr
                 -\fc && 1} \right)
\label{konedinv}
\end{equation}
so that
\begin{equation}
\fdf = \left[ {1\over 1 - (\fc)^2/\ff} \right]^{1/2}
\label{fdfoned}
\end{equation}
The limits of equation \ref{fdfoned}\ are interesting. If the foreground
component has a much different spectrum than the CMB, then their
shape vectors will be much different, and $\fc \rightarrow 0$.
In this case,  $\fdf$
goes to one. That is, a foreground component with a shape much different
than that of the CMB does not degrade the sensitivity of the experiment.
On the other hand a foreground component with a shape very close to
that of the CMB (so that $(\fc)^2\rightarrow \ff$) produces a very
large $\fdf$. To minimize the \fdf\ in a given experiment then,
one must measure at frequencies designed to
maximize the ``angle'' between the foreground spectrum
and the CMB spectrum.

In the case when there are two foreground sources
to project out, we define the three angles:
\begin{equation}
\cos\phi_{1} \equiv \hat F^{\rm cmb} \cdot \hat F^1
\qquad;\qquad
\cos\phi_{2} \equiv \hat F^{\rm cmb} \cdot \hat F^2
\qquad;\qquad
\cos\phi_{12} \equiv \hat F^1 \cdot \hat F^2.
\label{phitwod}
\end{equation}
Then, moving through the same steps as in the one foreground
case (but this time with the aid of {\it Mathematica}), one finds
that
\begin{equation}
\fdf = \left[ {\sin^2\phi_{12} \over
\sin^2\phi_{12} - \cos^2\phi_1 - \cos^2\phi_2
+ 2\cos\phi_1\cos\phi_2\cos\phi_{12}} \right]^{1/2}.
\label{fdftwod}
\end{equation}
Note again that in the limit that one of the
foregrounds is parallel to the CMB ($\cos\phi_1 =1$
or $\cos\phi_2=1$), the \fdf\ blows up as is expected.

\section{Applications}

\begin{figure}[htbp]
\centerline{
\hbox{
\psfig{figure=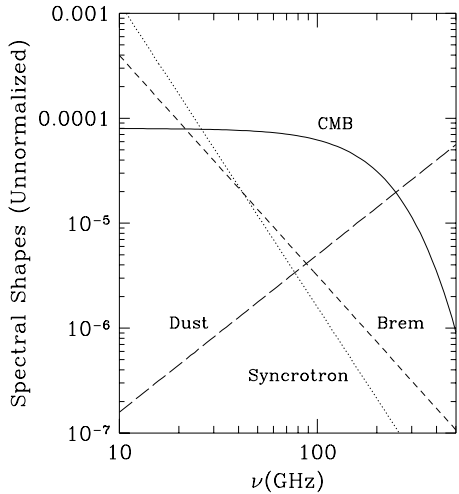,width=6.5cm}
}}
\caption[ ]{Unnormalized shapes of the different components in the
sky.}
\label{fig:figone}
\end{figure}

I now apply the formalism of the previous section to several
practical questions. To set the stage, consider figure
1 which shows the spectra of the three galactic
foregrounds of interest to us: synchrotron, bremsstrahlung,
and dust. The shape of the bremsstrahlung spectrum is
pretty well fixed by atomic physics. If we parametrize
a given shape by
\be{forshape}
\vec F^i_a \propto \nu^{p_i}_a
\ee
then $p_{\rm brem} \simeq -2.1$ with an uncertainty of a few
percent. The spectral index of synchrotron is much more
uncertain; typical estimates suggest that $p_{\rm sync}
= -2.9 \pm 0.2$. Finally the uncertainty in the spectral index
of dust is even more pronounced; in fact it is not even clear
if a fit along the lines of equation \ref{forshape}\ is adequate
to represent the complexities of dust. Nonetheless, a rough
estimate might give $p_{\rm dust} = 1.5 \pm 0.5$. Figure 1
illustrates these different shapes. Thus dust is expected
to dominate at high frequencies and the other components at low
frequencies.

\subsection{One Component: Synchrotron}

Let us start with the simplest possible example: one foreground
component, synchrotron, with spectral index assumed known.
This example, while crude, is not really
that unrealistic. At low frequencies dust can be safely ignored, and
bremmstrahlung typically comes in lower
than synchrotron. Further, as we will see in the next subsection, the
uncertainty in the spectral index introduces very little error.

According to equation \ref{finalvar}, the uncertainty in our
determination of the CMB temperature has only one piece
if the spectral shape of the foreground is known:
\be{onesync}
\sigma_\theta^2
= \fdf^2 \sigzt \rightarrow \fdf^2 {\sn^2\over \nch}
\ee
where the limit $\sigzt \rightarrow
\sn^2/\nch$ holds in the case of
 equal and uncorrelated noise in each channel
with variance $\sn$, as long as we are safely in the
the Rayleigh-Jeans limit.
 In this simple example with only one
foreground to be projected out, we saw in equation
\ref{fdfoned} that $\fdf=1/\left(1 - (\fc)^2/\ff\right)$
The dot products and hence the $\fdf$
depends on the shape we assume for synchrotron emission
(here I will assume $p_{\rm sync} = -2.9$) and also on the placement
of the frequency channels.

What is the optimal placement of frequency channels? And how many
are needed? Let us first consider two frequency channels. For simplicity
I will assume that they are centered about $\nu = 40$ GHz. Figure
2 shows the \fdf\ as a function of the difference $\nu_{\rm high}
-\nu_{\rm low}$. For example $\nu_{\rm high}
-\nu_{\rm low} = 10$ GHz indicates two channels placed at $35$
and $45$ GHz. The \fdf\ in that case is a little over three: the signal
to noise is degraded by this factor. For very small frequency differences,
it is difficult to disentangle the CMB component from synchrotron; hence
the $\fdf$ factor is high. If the frequencies can be spread far apart,
separating CMB from synchrotron becomes easier and the \fdf\ decreases
accrordingly. In the limit of very large frequency difference, the \fdf\
asymptotes to:
\be{asymptote}
\lim_{\Delta \nu \rightarrow \infty} \fdf = \sqrt{ { \nch\over\nch-\nf }}
\label{fdfass}
\ee
in this case $\sqrt{2}$. To see why, note that in the absence of foregrounds,
the additional information from all the channels beats down the
noise by a factor of $1/\sqrt{\nch}$; this is the factor explicitly
present in equation \ref{onesync}. When a foreground component is present,
at least one of the channels must be used to determine the foreground
amplitude. Thus even in the ideal case, when the foreground component
can be easily distinguished from the CMB, there is still one fewer
channel with which to measure the CMB. Hence, the true noise is now
down by a factor of $1/\sqrt{\nch-1}$. And on it goes as more foreground
amplitudes must be separated. Note that these arguments are only valid in
the Rayleigh-Jeans limit; For higher frequencies, the
limits in equation
\ref{onesync} and equation \ref{fdfass}\
no longer apply.

\begin{figure}[htbp]
\centerline{
\hbox{
\psfig{figure=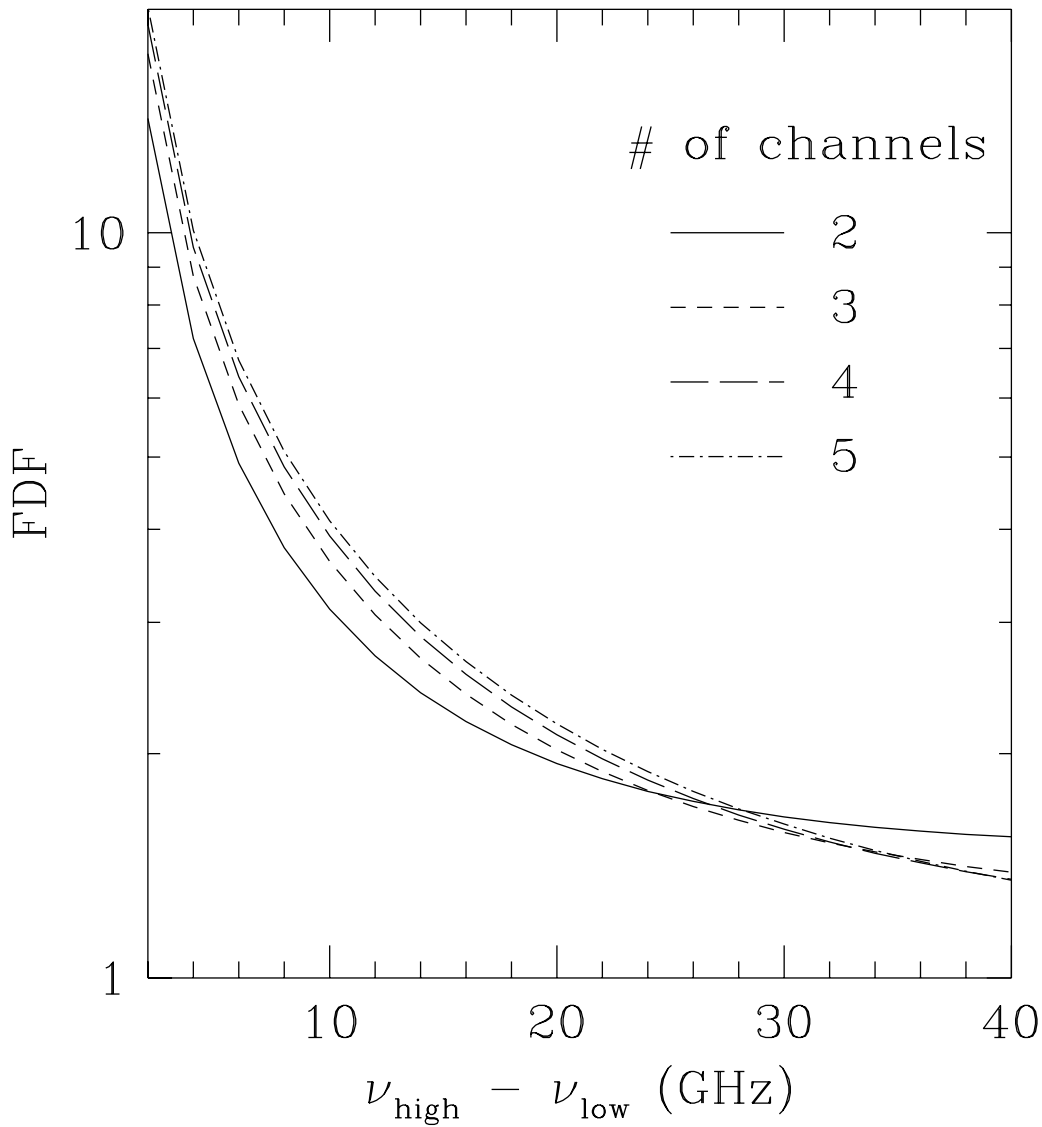,width=15cm}
}}
\caption[ ]{FDF as a function of the difference between the highest and
lowest frequencies in an experiment when synchrotron with
assumed index $-2.9$ is fitted for. The extreme channels are centered
around $40$ GHz (so $(\nu_{\rm low} + \nu_{\rm high})/2 = 40$ GHz). The
curves with more than two channels have their frequencies equally spaced
between the two extremes. FDF is lowered -- hence the experiment
has the best discrimination against foregrounds -- when the frequency
spread is as large as possible.}
\label{fig:figtwo}
\end{figure}

\begin{figure}[htb]
\centerline{
\hbox{
\psfig{figure=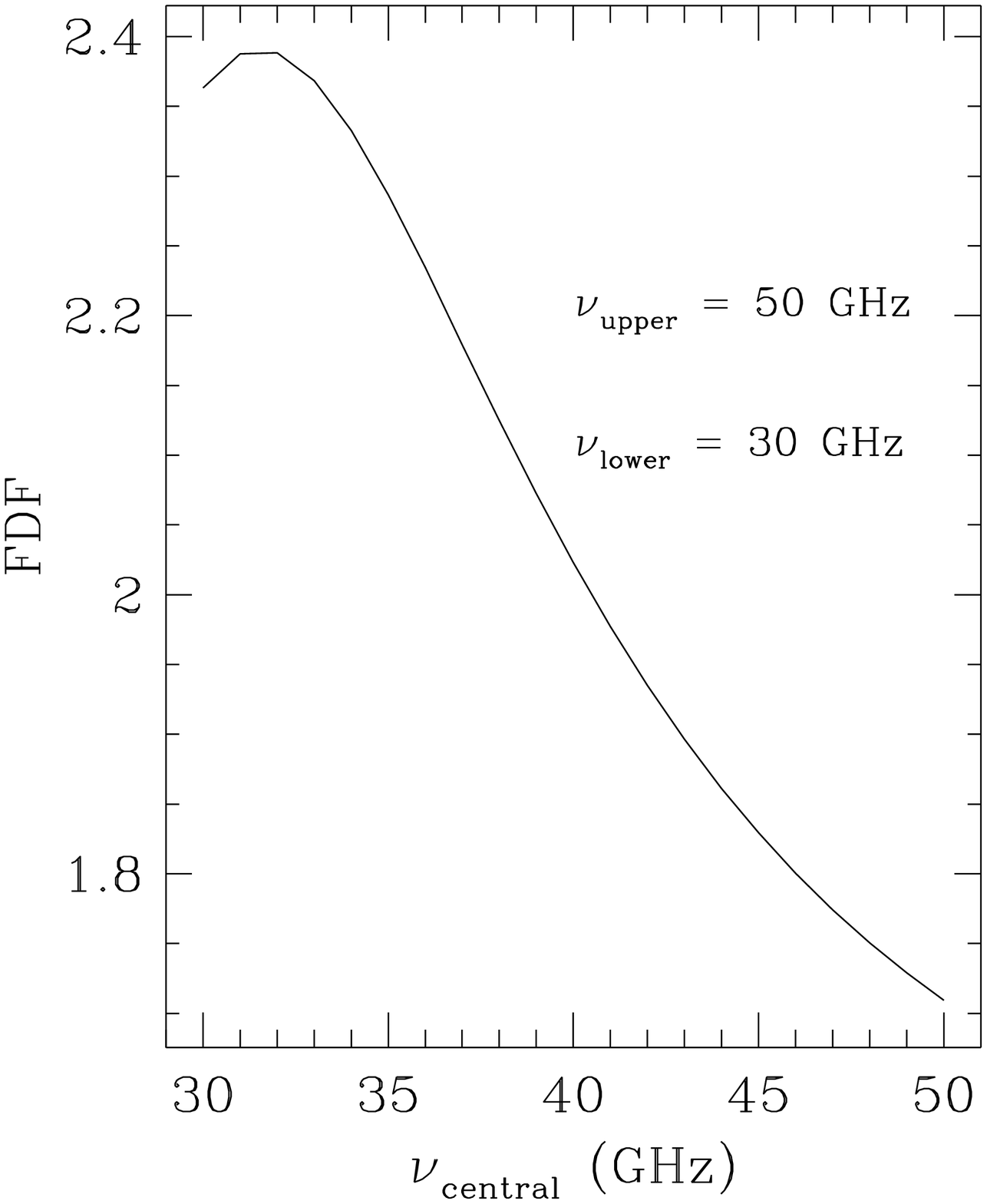,width=7.5cm}
}}
\caption[ ]{FDF as a function of the placement of the central frequency
in an experiment with three channels, the other two at $30$ and $50$
GHz. Again the foreground component is synchrotron with assumed index
$-2.9$. The optimal place to locate the third channel is at $50$ GHz, where
FDF is minimized.}
\label{fig:figthree}
\end{figure}

Now consider three frequency channels. It is clear that it is best
to get as large a spread in the frequencies as possible. But where best
to place the middle frequency channel? Figure 3 shows the \fdf\ as
a function of the frequency of the middle frequency channel when
$\nu_{\rm low} = 30$ GHz and $\nu_{\rm high} = 50$ GHz. Figure 3 shows
that the optimal place for the middle frequency channel is at $\nu=50$
GHz! At first, this is surprising, but it makes sense
upon further reflection:
the lowest channel is used to separate out the synchrotron
component. The other channels are best placed where they will get
the least contaminated by synchrotron; thus all other channels should
go as high in frequency as possible. Of course, this example is
somewhat artificial: when more than one foreground component is
projected out, it becomes important to space out the channels
more evenly. Nonetheless, I hope this simple example alerts
experimenters to the possibility that the best signal to noise may be
achieved with an unorthodox positioning of the frequency channels.
In this simple example, the \fdf\ varies from $2.3$ to $1.7$, i.e.
by roughly $30\%$, as one varies the placement of the middle channel.
So clever positioning of the intermediate channels could be an easy
way to increase the final signal to noise.

Figure 2 shows the \fdf\ for this case of three frequency channels
as a function of the difference between the highest and
lowest channel. In this graph, the middle channel is not placed in
the optimal position (usually at the highest frequency possible),
but all the channels are evenly spaced. (Thus the point
corresponding to $\nu_{\rm high}
-\nu_{\rm low} = 10$ GHz and $\nch=3$ has channels at
$\nu = 35,40,45$ GHz.)
It is interesting that, except for the largest frequency spreads,
adding extra channels does not really help in disentangling the
foregrounds. (This point was also made by Brandt et al. (1994).)
Certainly going beyond $\nch=3$ provides very little gain in this
simple case.

\subsection{Uncertain foreground shape}

We can generalize the discussion of the previous subsection
by accounting for the fact that the shape of the synchrotron
spectrum is not perfectly determined. If we allow for
this uncertainty, there arises a new term in the variance of
the estimator. Following equation \ref{sigfg}, we see that
\be{uncersyn}
\sigsh =
\vec T^{\rm sync} \cdot \left[ {K^{-1}}_{00} \hat F^{\rm cmb}
       + {K^{-1}}_{01} \vec F^{\rm sync}
\right]\ee
where again I emphasize that $T^{\rm sync}$ is the {\it true}
synchrotron temperature, with a spectral shape that
differs from the assumed one.
We will suppose that the true shape of the synchrotron
spectrum is still given by equation \ref{forshape}, but
with spectral index $p\ne -2.9$, the assumed index.
\begin{figure}[tbhp]
\centerline{
\hbox{
\psfig{figure=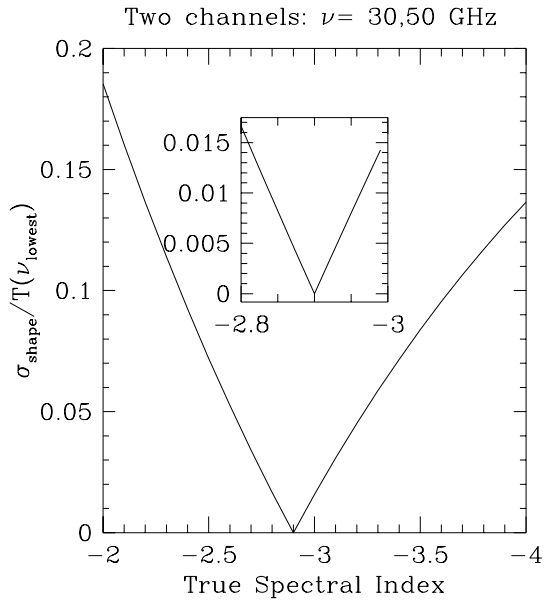,width=7.5cm}
}}
\caption[ ]{$\sigsh$:
the variance in the determination of the CMB temperature
due to uncertainty in the shape of the foreground spectrum being
fitted for. The assumed spectral index is $-2.9$; if the true index
is equal to this, then $\sigsh$ vanishes.}
\label{fig:figfour}
\end{figure}
Figure 4 shows the error induced by assuming the wrong
spectral index. For the kind of uncertainty typically
measured for synchrotron, $\Delta p \sim 0.2$, the
error induced is less than a few percent of the synchrotron
amplitude. Thus, if the synchrotron amplitude is
$40\mu$K, the uncertainty in the spectral index
contributes less than $1~\mu$K to the total error. This error
is very small and for reasonable noise values will be
much smaller than the $(\fdf)~ \sn/\sqrt{\nch}$ factor discussed
above.

It is instructive to understand why the uncertainty in the
spectral index leads to very small errors in the CMB
temperature determination. By projecting out the
$p=-2.9$ component, we are looking in the space perpendicular
to the shape vector defined by $p=-2.9$. But, the
vector defined by $p=-2.8$ is almost perfectly parallel to
the $p=-2.9$ shape vector. Thus, it has a very small
component in the
perpendicular space.
Unless the amplitude is extremely large, the perpendicular
component is negligible.

Figure 4 shows that, even when we project out only
synchrotron emission, we also succeed in eliminating
a large fraction of the bremmstrahlung (with
$p=-2.1$) as well. In this
example, only $15\%$ of the bremmstrahlung amplitude remains
after projecting out a $p=-2.9$ component. So
the simple projection of $p=-2.9$
is sufficient for all but the most sensitive experiments.

\begin{figure}[htbp]
\centerline{
\hbox{
\psfig{figure=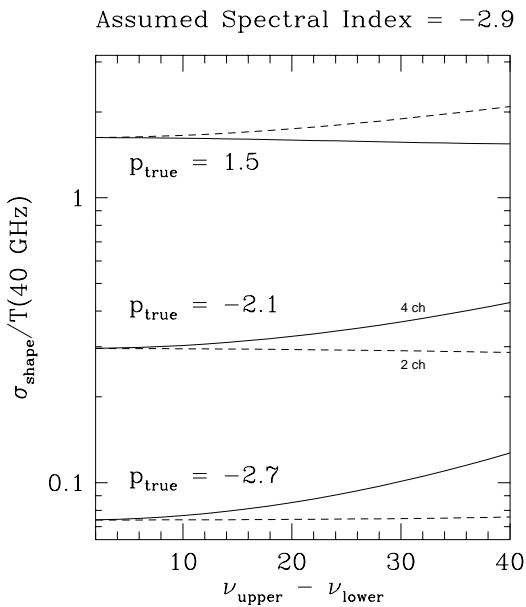,width=7.5cm}
}}
\caption[ ]{$\sigsh$
for different true
foreground shapes as a function of
the frequency range of an experiment. Solid
lines are for a 4 channel experiment (with equally spaced
frequencies); dashed lines for a 2 channel experiment.
In all cases, the range is centered around 40 GHz.}
\label{fig:figfive}
\end{figure}
Figure 5 shows how $\sigsh$ varies as the frequency
coverage changes. In this example, increasing the
frequency range always leads to an increase
in $\sigsh$. For some foregrounds, increasing the number
of channels also leads to an increase in $\sigsh$,
although this is not true for dust here. I have not
been able to figure out any general principles for
minimizing $\sigsh$; fortunately, it is simple
enough to deal with each case individually.

\subsection{Projecting out two components}

This subsection deals with an analysis question.
How best to analyze the data? In particular, should one
attempt to fit for several components or is it best to
fit for fewer components?  I will
argue that projecting out two components often will
lead to a larger variance than if one simply
projected out one component as in \S 3.1.

\begin{figure}[htbp]
\centerline{
\hbox{
\psfig{figure=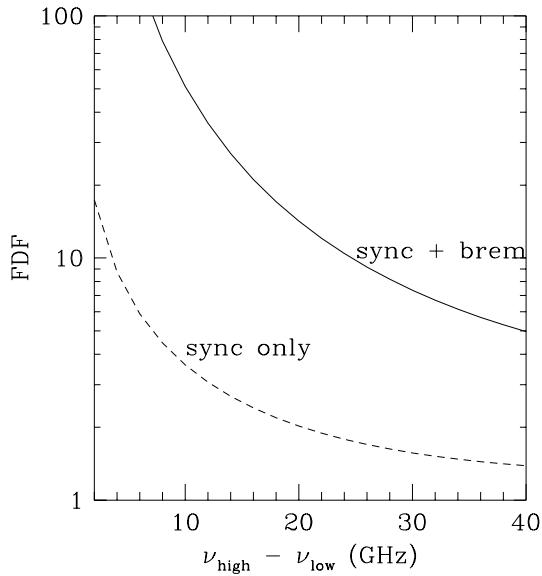,width=7.5cm}
}}
\caption[ ]{FDF as a function of frequency range in a three channel
experiment. Projecting out two components leads to a much larger
FDF.}
\label{fig:figsix}
\end{figure}

Figure 6 shows the \fdf\ for an experiment with three
frequency channels when both synchrotron [with index $-2.9$]
and bremmstrahlung are fitted for. For comparison,
also shown is the \fdf\ if only synchrotron was fitted for.
Again the central channel is at $\nu=40$ GHz. In all
cases, the \fdf\ is much higher if both components
are fitted for. For example,
with channels at $\nu =30,40,50$, figure 6 shows that
the ``one-component'' $\fdf=2$ while the ``two-component''
$\fdf=14$.

Let me pursue this example further. When is it
advantageous to project out two components? The total
variance in the two-component analysis is
\begin{equation}
\sigma_\theta^2\vert_{\rm two-component} =
(14)^2 \sigzt.
\label{twocom}
\end{equation}
There is also a small uncertainty due to the
unknown shape but this is very small so I neglect it.
In the one-component analysis, we
must include the shape uncertainty since the bremmstrahlung
amplitude is not projected out. Thus
\begin{equation}
\sigma_\theta^2\vert_{\rm one-component} =
(2)^2 \sigzt + (0.18)^2 \langle T^2_{\rm brem}(\nu=30 {\rm GHz})
\rangle.
\label{onecom}
\end{equation}
The coefficient $0.18$ in equation \ref{onecom}\ can be simply read
off of figure 4.  It becomes useful to analyze the
data by fitting for two components only when
$\sigma_\theta^2\vert_{\rm two-component} <
\sigma_\theta^2\vert_{\rm one-component}$. Using equations
\ref{twocom} and \ref{onecom}, we find that this occurs when
\begin{equation}
{\langle T^2_{\rm brem}(\nu=30 {\rm GHz}) \rangle^{1/2}
\over \sigz}
>
77.
\end{equation}
For even the most sensitive experiments, we do not expect
foreground amplitudes of this magnitude. So in this example,
it would be best to analyze the experiment by fitting for
only the synchrotron component.

One must pursue each example on a case by case basis. This
simple analytic technique should prove useful in deciding how
best to analyze the data. This simple example suggests that
fitting for fewer components leads to a smaller variance
in the CMB temperature; this agrees with the general point
made by Brandt et al. (1994). Now let us turn to a more
quantitative comparison with that work.

\subsection{Comparison with Brandt et al.}

The analytic techniques presented here can be compared
with the
Monte Carlos performed by Brandt et al. (1994). Here I focus on
one example of theirs, a seven channel experiment with
equally spaced frequencies between $25$ and $38$ GHz. I will
not describe their methodology in detail [please see their paper for a
lucid description of what they did]. For the purposes of
comparison, note that they were interested in the
same quantity I have been focusing on: the total variance
in the determination of the CMB temperature. I have called
it $\sigma_\theta$; they called it $E_{\rm RMS}$. Given the experimental
configuration and the average foreground levels, we can plot
this variance as a function of instrumental and atmospheric noise
per channel, $\sigma$ [in their notation roughly equivalent to
$\xi$]. Figure 7 shows a comparison of the analytic technique
and the Monte Carlos. The points are two different techniques that
they used to extract the CMB temperature. They
allow for free synchrotron amplitude [model $Q2$] and free synchrotron
and bremmstrahlung amplitudes [model $P3$]. This corresponds
to projecting out $\nf=1,2$ foregrounds respectively. The
curves show $\sigma_\theta$ with these two projections.
To get these curves I needed $\langle T_{\rm sync}^2 \rangle$
and $\langle T_{\rm brem}^2 \rangle$; I took the same values
they used in their Monte Carlos.

\begin{figure}[phtb]
\centerline{
\hbox{
\psfig{figure=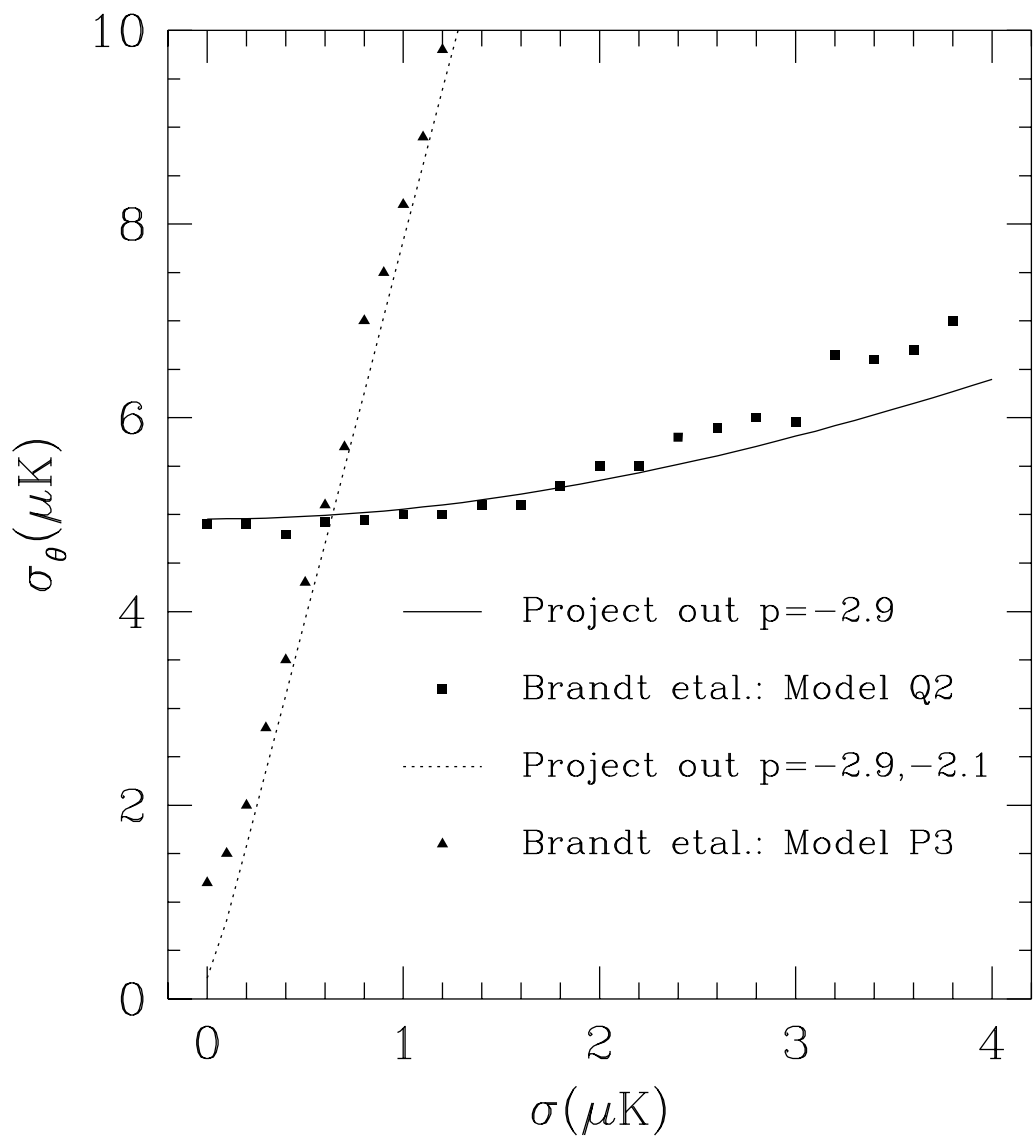,width=15cm}
}}
\caption[ ]{The variance in the determination of the CMB temperature
as a function of the noise per channel. The points denote two different
methods used by Brandt et al (1994) to extract the CMB temperature. The
lines are the variances one gets with the analytic formula of Eq. 1
fitting for one and two foreground components.}
\label{fig:figsev}
\end{figure}

The agreement is excellent and shows clearly
that the simple analytic technique adequately describes the
situation. The shapes of their curves now becomes obvious: at
low noise levels [small $\sigma$], $\sigsh$ dominates over the
\fdf-enhanced noise. Since $\sigsh$ is independent of noise,
the total variance is also independent of noise in this regime.
That is, at low $\sigma$, $\sigma_\theta$ is constant. In the
opposite limit, \fdf-enhanced noise dominates over $\sigsh$,
so the total variance increases linearly with $\sigma$.

One final point about our approaches. They presented
many other ``model''s for extracting out the CMB temperature.
For example, one of their models allowed the synchrotron index to be
a free parameter. Within the analytic framework presented
here, I cannot allow the shapes to be free parameters. However,
the variances using such models are much higher than the variances
in the models where only the amplitudes are allowed to vary. So
I would argue that the analytic technique cannot do everything but
it can do the things that are worth doing.

\subsection{Breaking up bands and noise correlations}

We have seen in the previous sections that adding more
frequency channels to an experiment is not necessarily
a good thing.
For, intermediate channels are not as effective in
separating out different spectra; a longer lever arm with
good sensitivity at both ends is often
preferable. In this section
I focus on another possible danger of splitting
up bands. Often when a given frequency band is split up,
the noise in the new channels is correlated. How does
correlated noise impact on the decision to split up
bands? Here, I address this question in the context of
a simple example.

Consider an experiment with two channels in the Rayleigh-
Jeans regime, say $\nu=15,45$ GHz. The noise in each channel
is assumed to have variance $\sigma$, so in the absence
of foregrounds, the variance in the determination of
the CMB temperature would be $\sigma/\sqrt{2}$. If
we wish to project out synchrotron emission (with
assumed spectral index $-3$), then this
variance is increased by the $\fdf$. In this case a
simple computation yields
$\fdf=1.51$, so the total variance in the experiment
is
\begin{equation}
\sigma_{\theta,2}
= 1.07 \sigma
\end{equation}
where the subscript ${}_2$ denotes the number of channels.
Is it worthwhile to add two new frequency channels
at $\nu=25,35$ GHz? I will asume that in so doing,
the noise in each channel increases by $\sqrt{2}$, so
that in the absence of foregrounds, the variance would
still be $(\sqrt{2}\sigma)/\sqrt{4} = \sigma/\sqrt{2}$.
If there were no correlations introduced between the
different channels, then we could do a simple
calculation and find that
$\fdf=1.33$. Thus the total variance in this $4-$channel
case is only $0.94\sigma$, smaller than in the two channel
case and perhaps worth the effort.

However, if correlations amongst the different channels
are introduced, then the calculation becomes slightly
less trivial. Here I carry out the calculation in this
correlated case for several reasons. First, this will
give us a sense of whether or not it is important
to avoid correlations. But more importantly, I hope that
this provides yet another example of how useful
the formalism of \S 2 can be when it comes to analyzing
specific problems.

For simplicity I will assume that the two lowest channels
are correlated as are the two highest channels, so the
new noise correlation matrix is
\begin{equation}
C = 2\sigma^2 \left(
\begin{array}{clcr}
1 & \epsilon & 0 & 0\\
\epsilon & 1 & 0 & 0\\
0 & 0 & 1 &\epsilon \\
0 & 0 & \epsilon & 1
\end{array}
\right).
\end{equation}
For the calculation we will need the inverse of $C$. A short
calculation shows that
\begin{equation}
C^{-1} = {1\over 2\sigma^2(1-\epsilon^2)} \left(
\begin{array}{clcr}
1 & -\epsilon & 0 & 0\\
-\epsilon & 1 & 0 & 0\\
0 & 0 & 1 &-\epsilon \\
0 & 0 & -\epsilon & 1
\end{array}
\right).
\end{equation}
We can now immediately calculate the variance in the absence
of foregrounds:
\begin{eqnarray}
\sigma_\theta^{(0)^2}
&=& {1\over \sum_{ab} {C^{-1}}_{ab}} \nonumber\\
&=& {\sigma^2 (1+\epsilon)\over 2}
\end{eqnarray}
where in the first line I have assumed that we are deep
enough into the Rayleigh-Jeans to set $\hat F^0_a = 1$.
Thus, the variance in such an experiment -- in the absence
of foregrounds -- increases due to correlations
by a factor of $\sqrt{1+\epsilon}$. This is
a very simple way of saying what I and my
collaborators tried to illustrate in
Dodelson, Kosowsky, and Myers (1995). Now let us include
the effects of foregrounds. Our standard formula gives
\begin{equation}
\fdf^2 = {1\over 1 - (\vec F^1\cdot\hat F^0)^2/
(\vec F^1\cdot\vec F^1)}
\label{fdfcor}
\end{equation}
so we need to calculate the two dot products. The only
complication here is that we need to account for
the non-diagonal structure of $C$. Thus,
\begin{eqnarray}
\vec F^1\cdot\hat F^0 &=&
\sigma_\theta^{(0)^2} \sum_{ab}
	{C^{-1}}_{ab} F^1_b \nonumber\\
&=& \sigma_\theta^{(0)^2} {\sum_a F^1_a
\over 2\sigma^2 (1+\epsilon)}
\end{eqnarray}
and
\begin{eqnarray}
\vec F^1\cdot\vec F^1 &=&
\sigma_\theta^{(0)^2} {\sum_a (F^1_a)^2
-2\epsilon (F^1_1 F^1_2 + F^1_3 F^1_4 )
\over 2\sigma^2 (1-\epsilon^2)}.
\end{eqnarray}
With these expressions for the dot products we can now
evaluate the \fdf\ with equation \ref{fdfcor}.

\begin{figure}[htbp]
\centerline{
\hbox{
\psfig{figure=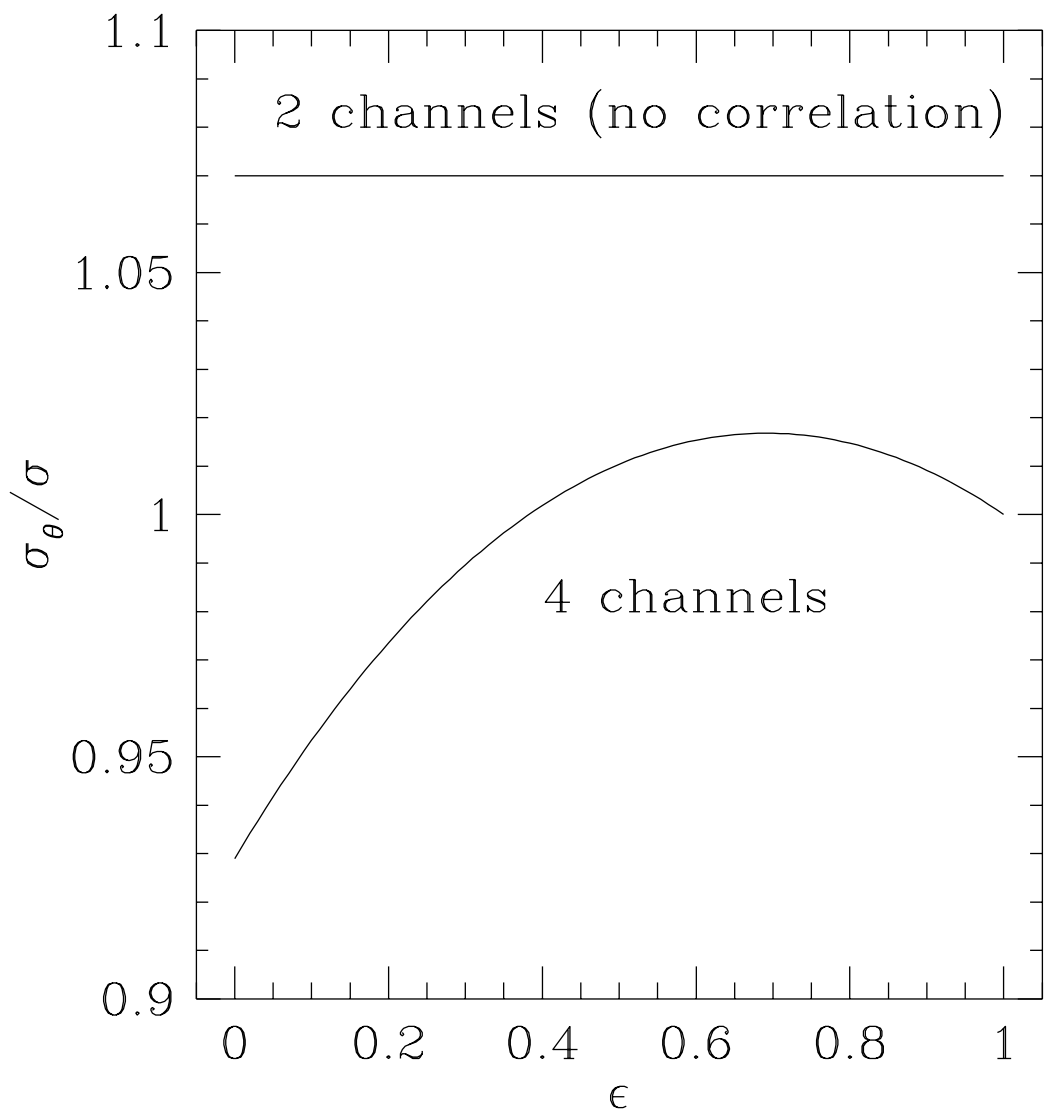,width=15cm}
}}
\caption[ ]{The variance in the determination of the CMB
temperature as a function of the correlation amongst the
different frequency channels. This is to be
compared with the horizontal line, the variance
in the two channel case when there is no correlation.
Since the four channel curve is lower than the horizontal
line, it would always be advantageous to add the extra
channels in this case even if correlations were introduced.}
\label{fig:figeig}
\end{figure}

Figure 8 shows the variance in the determination
of the CMB temperature as a function of the correlation
between the channels when the synchrotron is fitted for.
Apparently, increased correlation does
not significantly increase the variance. So,at least in
this example, noise correlation should not deter
experimenters from adding new channels.

\subsection{Current Experiments}

\def\reff{\par\noindent\hangafter=1\hangindent=1cm}
To get a feel for how well current experiments are doing at
separating out foregrounds, I compiled Table 1. For each
experiment, the \fdf\ is computed for a given spectral index.
For example the \fdf\ for COBE fitting for bremmstrahlung is
$1.75$. Also shown is the uncertainty due to the shape. Again for
COBE, if bremmstrahlung is fit for, then a dust
component [with index $1.5$]
contributes an uncertainty $\sigsh = 5.32 \langle T_{\rm lowest}^2
\rangle^{1/2}$. For COBE, the lowest frequency channel is at $31$ GHz.
At this frequency, one expects an rms dust antenna temperature of order
a few $\mu$K, so -- in the absence of any other maps -- the uncertainty
due to dust would be of order $10 \mu$K. [This is not intended to
be a rigorous estimate of the uncertainty due to dust, just a guide
to reading the table. COBE has access to -- and used much other information
to get a handle on dust. See for example Bennett et al. (1994).]

\begin{table}[p]
\caption[]{FDF's for selected experiments.}
\label{table:first}
\begin{center}
\mbox{
\begin{tabular}{|l||c|c|c|}
\hline
Experiment  &  Assumed Index  &  FDF   &
Foreground with $p=\qquad\rightarrow$ $\sigma_{\rm shape}/T_{\rm lowest}$  \\
\hline
\hline
COBE$^a$  &  -2.1  &  1.75  &  1.5  $\rightarrow$ 5.32 \\
FIRS$^b$  & 1.5  &  1.02  &  2 $\rightarrow$ 2.45 \\
MAX3$^c$  &  -2.1  &  11.5  &  1.5 $\rightarrow$ 73.7 \\
MAX4$^d$  & -2.1  &  4.09  &  1.5  $\rightarrow$ 12.4 \\
MAX3  & 1.5  &  1.12  &  -2.1  $\rightarrow$ 2.09 \\
MAX4  & 1.5  &  1.06  &  -2.1 $\rightarrow$ 1.16 \\
MAX3  &  1.5  &  1.12  &  2 $\rightarrow$ 1.34 \\
MAX4  &  1.5  &  1.06  &  2 $\rightarrow$ 1.95 \\
MSAM1$^e$  &  1.5  &  1.02  &  2 $\rightarrow$ 2.48 \\
SK93$^f$  &  -2.9  &  4.48  & -2.1  $\rightarrow$ .234 \\
SK94$^g$  & -2.9  &  2.35  &  -2.1 $\rightarrow$ .180 \\
SP91$^h$ &  -2.9  &  4.00  &  -2.1  $\rightarrow$ .225 \\
SP94$^i$ &  -2.9  &  2.40  &  -2.1 $\rightarrow$ .179 \\
Tenerife$^j$  & -2.9  &  1.54  &  -2.1 $\rightarrow$ .094  \\ \hline
Satellite  &  -2.9,1.5  &  2.97  &  -2.1 $\rightarrow$ .169  \\
\hline
\end{tabular}
}
\end{center}
\reff $^a$ Bennett et al. (1994)
\reff $^b$ Ganga et al. (1994)
\reff $^c$ Meinhold et al. (1993)
\reff $^d$ Clapp et al. (1995)
\reff $^e$ Cheng et al. (1994)
\reff $^f$ Wollack et al. (1993)
\reff $^g$ Netterfield et al. (1994)
\reff $^h$ Gaier et al. (1992)
\reff $^i$ Gundersen et al. (1994)
\reff $^j$ Hancock et al. (1994)
\end{table}

A cursory look at some of the other experiments in Table 1 shows that typical
\fdf's are of order $1-4$. Bolometer experiments like MAX and MSAM do very
well at projecting out dust. [Note though that MAX3 in particular could
not distinguish well between CMB and bremmstrahlung.] The HEMT
experiments do not have large frequency coverage, so they discriminate
less well than the high frequency experiments. However,
the recent modifications
to the South Pole and the Saskatoon [additions of higher frequency channels]
have significantly reduced their \fdf's.

The last line of Table 1 presents the
 \fdf\ for a hypothetical
experiment with equally spaced frequencies between $30$ and $120$ GHz
(this case was also analyzed by Brandt et al. (1994)). Analyzing by
projecting out two components leads to a variance
squared
equal to
$(2.97\times \sigma/\sqrt{7})^2 + (.169 T_{\rm brem}(30 {\rm GHz}))^2$,
where $\sigma$ is the noise in each channel.

\newpage

\section{Conclusions}

This paper has introduced an analytic technique
that can be used to help design an experiment and
to analyze data. The main result coming out of this
analytic treatment is that the variance in the determination
of the CMB temperature has two components. First, there
is a component proportional to noise; due to
fitting for foregrounds, noise is amplified by the
\fdf. Second, there is a component $\sigsh$ proportional
to the foreground amplitudes. This component vanishes
if the foreground spectra are known, but is non-zero
if there is some uncertainty in the shapes. This simple
model of CMB extraction was shown in \S 3.4 to
reproduce the Monte Carlo results of Brandt et al (1994)
very accurately.

I think the most useful thing to
emerge is the technique itself, which is easy to understand
and implement. Any given experiment will have its own set
of complications, so it is dangerous to make general
conclusions about the ``best'' set of frequency channels.
Nonetheless, on the basis of the work in \S 3, there
are several general principles that should be considered
in any experimental plan/analysis.

\begin{itemize}

\item {\it
A wide range of frequencies does best at minimizing
the FDF.} In general this would lead one to go with
large frequency ranges. Indeed, I would argue that
experiments with bolometers have been more
successful to date at extracting the CMB since they allow
a larger frequency range. However, one can envision
circumstances where increasing the range is not beneficial
\footnote{I am grateful to Steve Levin for emphasizing
this point to me.}. In particular, as shown in \S 3.2,
increasing the frequency range often leads to a larger
$\sigsh$. This effect can be even more dramatic if the
new frequencies are more sensitive to a different
foreground component [e.g. a $120$ GHz channel added to
a low frequency experiment would be more sensitive
to dust].

\item Equal spacing of the intermediate  channels is not
always the optimal way to go. Furthermore, adding more
intermediate channels is also not necessarily beneficial.
However, it does not appear -- at least from the example
studied in \S 3.5 -- that noise correlations amongst
different channels should be a deterrent in this regard.

\item In terms of analysis, in agreement with the results
of Brandt et al. (1994), I found in \S 3.3
that it is best to fit
for as few components as possible. A cursory glimpse at
present experiments suggests that their signal to noise
is degraded due to fregrounds
by a factor ranging from one to four, with
bolometers at the low end and HEMPTs at the high end.
A satellite experiment with frequencies ranging from
$30$ to $130$ GHz would see its signal to noise degraded
by about three.

\end{itemize}

\section*{Acknowledgments}
Many people on different experimental groups
have been thinking about similar problems to the ones
discussed here. I thank them for discussions. Especially
helpful were Gary Hinshaw, Lloyd Knox, and Steve Levin.
I am also grateful to
Arthur Kosowsky and Albert Stebbins,
my collaborators on previous aspects of foreground separation.
This work was supported in part by the DOE (at Chicago and
Fermilab) and the NASA (at Fermilab through grant NAG 5-2788).

\appendix

\section{Derivation of Variances}

We want to solve
equation \ref{minvar} for the free parameters, the
amplitudes $\theta^i$. This minimization requirement is satisfied when
\begin{equation}
 \vec F^j \cdot \left( \vec T - \sum_{i=0}^{\nf} \vec F^i \theta^i
\right) = 0 .
\label{mineq}
\end{equation}
Using the definition of $K$ in equation \ref{defk}\ leads to
\begin{equation}
\sum_{i=0}^{\nf} K_{ji} \theta^i = \vec F^j \cdot \vec\tobs.
\label{eqinter}
\end{equation}
Multiplying by $K^{-1}$ and summing over $j$ leads to
\begin{equation}
\theta^i = \sum_{j=0}^{\nf} {K^{-1}}_{ij} \vec F^j \cdot \vec\tobs.
\label{eqesti}
\end{equation}
The $i=0$ component of this equation is the estimator for
the CMB temperature presented in equation \ref{eqest}.

Now we want to calculate the variance of the
estimator for the CMB temperature. Start from equation
\ref{defvar} and use \ref{eqest}\ for $\theta^0$ and
equation \ref{obssig}\ for $\vec T$. Then,
\begin{equation}
\sigma_\theta^2 =
 \langle\Bigg( \sum_{j=0}^{\nf} K^{-1}_{0j} \vec F^j \cdot
\bigg[\sum_{i=0}^{\nf} \vec T^i  + \vec N \bigg]
 - t^{0} \Bigg)^2 \rangle.
\label{intvar}
\end{equation}
Consider the $i=0$ term here.  This is
\begin{equation}
\sum_j K^{-1}_{0j} \vec F^j \cdot \hat F^0 t^{0}
= \sum_j K^{-1}_{0j} K_{j0} t^{0} = t^{0}
\label{intstep}
\end{equation}
so this part of the sum exactly cancels the $t^{0}$ term
in equation \ref{intvar}. Thus we are left with
\begin{equation}
\sigma_\theta^2 =
 \langle\Bigg( \sum_{j=0}^{\nf} K^{-1}_{0j} \vec F^j \cdot
\bigg[\sum_{i=1}^{\nf} \vec T^i  + \vec N \bigg] \Bigg)^2 \rangle.
\label{intivar}
\end{equation}
The first term in square brackets is exactly $\sigsh$ in equation \ref{sigfg}.
The noise term is slightly more complicated. It is
\begin{eqnarray}
\langle\Bigg( \sum_{j=0}^{\nf} K^{-1}_{0j} \vec F^j \cdot
\vec N\Bigg)^2 \rangle
&=&
\sum_{j=0}^{\nf} K^{-1}_{0j}  \sigzt \sum_{ab} F^j_a  {C^{-1}}_{ab}
\sum_{j'=0}^{\nf} K^{-1}_{0j'} \sigzt \sum_{a'b'} F^{j'}_{a'} {C^{-1}}_{a'b'}
\langle N_b N_{b'}\rangle \nonumber \\
&=&
\sigma_\theta^{(0)^4} \sum_{jj'} K^{-1}_{0j} K^{-1}_{0j'}
\sum_{aa'bb'} F^j_a F^{j'}_{a'} {C^{-1}}_{ab} {C^{-1}}_{a'b'}  {C}_{bb'}
\nonumber \\
&=&
\sigzt \sum_{jj'} K^{-1}_{0j} K^{-1}_{0j'}
\left( \sigzt \sum_{aa'} F^j_a {C^{-1}}_{aa'} F^{j'}_{a'} \right)
\label{mulvar}
\end{eqnarray}
In going from the first to the second line here
I have used equation \ref{noiseapp}. The term
in parentheses on the last line
in equation \ref{mulvar}\ is by definition
equal to $\vec F^j \cdot \vec F^{j'} \equiv K_{jj'}$.
This contracts with one of the $K^{-1}$'s to give
$\delta_{0j}$. Thus all that is left of the sum
is ${K^{-1}}_{00}$. This corresponds to the \fdf\
in equation \ref{deffdf}.

\section{Marginalization}

This appendix presents what appears to be another way
to extract the CMB signal. For a long time I thought
that this estimator was better than the one presented
in the body of the paper. I even gave a talk or
two explaining that this estimator was preferable
to any other. This is not true. Both
estimators are identical. In this appendix, I first
present the other method and then prove that
both estimators, although they look completely
different, are in fact identical. Albert Stebbins
and I in Dodelson $\&$ Stebbins (1994) wrote about
marginalization and Dodelson $\&$ Kosowsky (1995) describes
some more marginalization work. All of this is now
shown to be equivalent to the best fit technique used for
example by Cheng et al. (1994) to analyse the MSAM data.

The idea is to project out all the foreground sources.
The $\nf$
foregrounds span an $\nf-$ dimensional subspace of
$\Re^{\nch}$; call this subspace $\ssp$. All foreground
contributions to the signal live in $\ssp$.
The {\it orthogonal complement} of $\ssp$, $\ssp_\bot$,
is the space which contains all vectors orthogonal
to the foregrounds. Thus any vector in $\ssp_\bot$
is completely independent of foregrounds. What we need to
do, therefore, is project the observed temperature
vector on to $\ssp_\bot$. This projected temperature
will be independent of any foregrounds and hence will
provide an unbiased estimate of the CMB temperature.
To project on to $\ssp_\bot$, we first need a set of
basis vectors in $\ssp_\bot$. Let us call these
\begin{equation}
\hat \zr \qquad r = 1, \ldots, \nch-\nf .\end{equation}
These basis vectors are chosen to be orthonormal, so
they are perpendicular to each other and they have unit
norm:
\be{zone} \hat \zr \cdot \hat \zs =  \delta_{rs}
.\ee
Of course there are $\nch-\nf$ of these vectors since they
span the $\nch-\nf$ dimensional space $\ssp_\bot$.
Finally, by the definition of $\ssp_\bot$, the vectors
$\zr$ must satisfy
\be{ztwo}
\hat \zr \cdot \vec F^i = 0 \qquad r=1,\ldots,
\nch-\nf\quad;\quad i=1,\ldots,\nf
.\ee
With the basis vectors $\zr$, we can form the projection
operators which project any vector in the full $\nch$
dimensional space on to $\ssp_\bot$, the space independent
of foregrounds. For an arbirtary vector $\vec x$ in the full space,
\begin{equation}
\vec x_\bot \equiv \sum_{r=1}^{\nch-\nf} \hat \zr \left(\hat \zr
\cdot \vec x\right)
\end{equation}
is the projection on to $\ssp_\bot$. Thus $\vec x_\bot$ is
independent of any foregrounds.

We are now in a position to get the marginalization
estimate for
the CMB temperature. First we project the observed
temperature on to the space independent of foregrounds.
Then we find the CMB component of this
projected temperature. The estimator is therefore
\be{eqestmar}
\theta' \equiv {\hat\fcmb\cdot \perp\tobs
\over \hat\fcmb\cdot \hat\fcmb_\bot }
= \sum_{r=1}^{\nch-\nf}\
\left( \hat\fcmb\cdot\hat\zr\right)
\ \left(\hat\zr\cdot \vec\tobs\right)
\bigg/\sum_{r=1}^{\nch-\nf}\
\left( \hat\fcmb\cdot\hat\zr\right)^2
\ee
The denominator in equation \ref{eqestmar}\ is simply to
get the normalization right. [Thus when $\vec T = \hat F^0 t^0$,
the estimator $\theta'$ will give $t^0$].

This estimator looks [to me] completely different from the estimator
in equation \ref{eqest}. I now show that the two are
equivalent. Let me write $\theta=\vec a \cdot \vec T$
and $\theta'=\vec {a'} \cdot \vec T$ so
\begin{eqnarray}
\vec a &\equiv& \sum_{j=0}^{\nf} {K^{-1}}_{0j} \vec F^j
\nonumber\\
\vec{a'} &\equiv& {\hat\fcmb_\bot
\over \hat\fcmb\cdot \hat\fcmb_\bot }
\end{eqnarray}
If I can show that these two vectors are equivalent, then
I have shown that the two estimators $\theta$ and $\theta'$
are also equivalent.
One way to do this is to pick a basis
which spans the full $\nch$ dimensional space and show that
for each basis vector $\vec b$, $\vec a\cdot \vec b =
\vec {a'}\cdot \vec b$. As a basis consider the $\nf$ vectors
$\vec F^j (j>0)$ together with the $\nch-\nf$ unit vectors
$\hat\zr$. It is easy to check that
$\vec a\cdot \vec F^j = \vec {a'}\cdot \vec F^j = 0$ for all $j$.
Of course this is the way the estimators were constructed, to be
independent of foregrounds. Now I will show that
$\vec a\cdot \hat\zr = \vec {a'}\cdot \hat\zr$ for all $r$.
\begin{eqnarray}
\vec a\cdot \hat\zr &=&  (K^{-1})_{00} \hat F^0 \cdot \hat\zr
\nonumber\\
\vec {a'}\cdot \hat\zr  &=& {1
\over \hat\fcmb\cdot \hat\fcmb_\bot }~
\hat F^0 \cdot \hat\zr
\end{eqnarray}
Thus to show that the two estimators are identical, I need only show
that
\begin{equation}
{1
\over \hat\fcmb\cdot \hat\fcmb_\bot }
= (K^{-1})_{00}
\label{qedpr}
\end{equation}

To prove equation \ref{qedpr}\ let us calculate
\begin{equation}
\vec a\cdot \vec a = \sum_{jj'} (K^{-1})_{0j} (K^{-1})_{0j'}
\vec F^j\cdot\vec F^{j'} =
(K^{-1})_{00}
\label{asqo}
\end{equation}
since $\vec F^j\cdot\vec F^{j'} \equiv K_{jj'}$.
Since $\vec a$ is
perpendicular to all the foregrounds it lies in $\ssp_\bot$.
Thus it can be written as
\begin{equation}
\vec a = \sum_{r} \hat\zr \left(\vec a \cdot \hat\zr\right).
\end{equation}
Therefore, another way of writing the dot product in equation
\ref{asqo}\ is
\begin{eqnarray}
\vec a \cdot \vec a &=& \sum_{rr'} \hat\zr \left(\vec a \cdot \hat\zr\right)
\cdot \hat z^{(r')} \left(\vec a \cdot \hat z^{(r')}\right)
\nonumber\\
&=& \sum_{r} \left(\vec a \cdot \hat\zr\right)^2
\nonumber\\
&=&
(K^{-1}_{00})^2 \sum_{r} \left(\hat F^0 \cdot \hat\zr\right)^2
\end{eqnarray}
Using equation \ref{asqo}, we can now equate
\begin{equation}
{1\over  \sum_{r} \left(\hat F^0 \cdot \hat\zr\right)^2}
= (K^{-1})_{00}.
\end{equation}
But the left hand side here is equal to the left hand side of
equation \ref{qedpr}\ by definition. So the identity is proven.

\newpage
\begin{center}
{\bf References}
\end{center}

\reff C. L. Bennett et al. 1994, ApJ, 436, 423
\reff  W.N.Brandt, C.R. Lawrence, A.C.S. Readhead,
J.N. Pakianathan, \& T.M. Fiola, 1994, ApJ, 424, 1
\reff E. S. Cheng et al. 1994, ApJ, 422, L37
\reff A. C. Clapp et al. 1994, astro-ph/9404072
\reff L. Danese, L. Toffolatti, A. Frenceschini, M.
Bersanelli, \& N. Mandolesi 1995, astro-ph/9501043
\reff S. Dodelson \& A. Stebbins 1994, ApJ, 433, 440
\reff S. Dodelson \& A. Kosowsky 1995, PhysRevLett,
75, 604
\reff S. Dodelson, A.Kosowsky, \& S. T. Myers 1995,
ApJ, 440, L37
\reff T. Gaier, J. Schuster, J. Gundersen, T. Koch,
M. Seiffert, P. Meinhold, \& P. Lubin 1992, ApJ, 398, L1
\reff K. Ganga, L. Page, E. Cheng, \& S. Meyer 1994,
astro-ph/9404009
\reff J. Gundersen et al. 1994, astro-ph/9401220
\reff S. Hancock, R. D. Davies, A. N. Lasenby,
C. M. Gutie'rrez de la Cruz, R. A. Watson,
R. Rebolo, \& J. E. Beckman 1994, Nature, 367, 333
\reff P. Meinhold et al. 1993, ApJ, 409, L1
\reff C. B. Netterfield, N. C. Jarosik, L. A. Page,
D. Wilkinson, \& E. J. Wollack 1994, astro-ph/9411035
\reff Neugebauer et al. 1984, ApJ, 278, L1.
\reff M. Tegmark \& G. Efstathiou 1995, astro-ph/9507009
\reff L. Toffolatti et al. 1994, astro-ph/ 9410037
\reff E. J. Wollack, N. C. Jarosik, C. B. Netterfield,
L. A. Page, \& D. Wilkinson 1993, ApJ, 419, L49

\end{document}